\documentclass[a4paper,12pt]{article}

\begin{document}

\title{Local Indistinguishability and Possibility of Hiding cbits in Activable Bound Entangled States}

\author{Indrani Chattopadhyay\thanks{ichattopadhyay@yahoo.co.in}
and Debasis Sarkar\thanks{dsappmath@caluniv.ac.in}\\
Department of Applied Mathematics, University of Calcutta,\\
92, A.P.C. Road, Kolkata- 700009, India} \maketitle
\begin{abstract}
In this letter we prove local indistinguishability of four
orthogonal activable bound entangled states shared among even number
of parties. All reduced density matrices of such states are
maximally mixed. We further proceed to establish a multipartite
quantum data hiding scheme on those states and explore its power and
limitations.

PACS number(s): 03.67.Hk, 03.65.Ud, 03.65.Ta, 03.67.-a, 89.70+c.

Keywords: Local Indistinguishability, Data hiding, LOCC,
Entanglement.
\end{abstract}

\maketitle

Keeping a data secret by sharing it among some parties is an
important task in quantum information processing
{\cite{classicalsecret, quantumsecret}}. Secrecy of a data is
defined in two ways. Firstly against the attack of an eavesdropper
{\cite{com}} and secondly against the cheating attempts of the
parties sharing the data where the data is kept secret from the
parties themselves. A well known task in classical secret sharing is
to prepare a key, which is being distributed among some parties so
that to unlock the secret, i.e., to know the key, some parties (the
number of such parties can be pre-assigned) have to contribute their
shared parts {\cite{classicalsecret}}. Instead of classical key, if
quantum states are used to encode classical data, then we find two
different directions of research. In Quantum Secret Sharing the
hidden data can be explored by some of the parties concerned, by
collective LOCC (i.e., Local operations with classical
communications) on their shared parts and the maximum number of
allowable cheating parties can be restricted in the time of
construction of the protocol {\cite{quantumsecret}}. Another
upcoming area of research is Quantum Data Hiding, where to reveal
the secret it is not sufficient to use LOCC even if an arbitrary
number of parties are involved in the cheating process. The study in
this direction was motivated by the discovery of `Quantum
Nonlocality Without Entanglement' {\cite{nwe}}, which establishes a
very strange phenomenon that there are sets of orthogonal product
states, not LOCC distinguishable.

Distinguishability of quantum states has immense importance in
quantum information processing. For perfect discrimination, the set
of states must be mutually orthogonal. However, for composite
systems the situation is quiet different. In such cases, it would be
preferred to restrict the set of allowable operations to be local in
nature (i.e., LOCC). It is really hard to distinguish locally a set
of quantum states (entangled or not) shared between a number of
parties situated at distant places. Rather it shows many
counter-intuitive results in quantum information theory. It is found
that some orthogonal product states are locally indistinguishable
\cite{nwe}. In contrast, there are orthogonal entangled states that
are locally distinguishable \cite{hardy, dist, dist1}. In this work,
we have concentrated on local indistinguishability of some
multipartite mixed entangled states. Multipartite entanglement is
difficult to detect and very hard to characterize perfectly. Only
some symmetric structures are available and usable in practical
senses. Such symmetries sometimes provide the system an immense
power to perform some otherwise impossible tasks. Here, we proceed
in a quite different way to negate all the possibility of
discriminating a set of four highly symmetric multipartite mixed
states, shared between an even number of parties each holding a
qubit system, by LOCC. It is proved that if such four states can be
discriminated even with a small probability by LOCC, where single
copy of one of them are given, then it is possible to distill out
some positive amount of entanglement by some local processes, from a
bound entangled state. This idea is quite similar in some senses,
with the one given by Ghosh \emph{et.al} \cite{ghosh}, to show the
local indistinguishability of the four Bell states. However, we have
considered here a general class of orthogonal mixed multipartite
bound entangled states shared between even number of parties.

The local indistinguishable character and some other properties of
our activable bound entangled states provide us the possibility of
hiding classical bits in quantum states. We consider here the task
of quantum data hiding to hide classical data in quantum states with
a much more secured scenario. In quantum data hiding, classical
information is kept secret in terms of quantum states shared among
some parties situated at distant locations. The involved parties
know which quantum state is used to encode which classical bit, but
do not know the actual state they are sharing. The security in such
schemes must guarantee the requirement that the parties can not
retrieve the secret by LOCC only. This imply, in a quantum data
hiding scheme the hiding states must be necessarily locally
indistinguishable. Such processes should necessarily require some
amount (which may be pre-assigned) of quantum communication
{\cite{locc}}, i.e., exchange of quantum information, to retrieve
the hidden information. That pre-assigned amount of quantum
communication defines the level of security of the hiding scheme.
Previous works {\cite{hide1, hide2}} suggest that the hiding states
may be chosen to be separable. In case of pure states, maintaining
the primary requirement of orthogonality and local
indistinguishability property, it is impossible to find suitable
pair of pure orthogonal entangled or separable states {\cite{hardy}}
to hide one cbit of information. In entanglement based hiding
schemes where the hiding states are taken to be entangled, it is
expected that the scheme may be broken by a finite amount of prior
entanglement shared between the unfaithful parties who may cheat
others and try to retrieve the data. The aim of such a hiding scheme
is to build a considerably high level of security with a minimum
number of faithful parties, required to maintain the secrecy. By
faithful parties we mean those who are not try to recover the hidden
data by exchange of quantum information. In any such scheme, the
hiding states are expected to have a highly symmetrical structure to
construct the security bound, independent of any permutation of
unfaithful parties. For that reason, only the number of unfaithful
parties is important to establish the security of the protocol.
Schemes are also proposed to encode quantum data in terms of qubits
into hiding states and in bipartite case, it is found that hiding
two classical bits is equivalent of hiding a qubit in a similar
scenario {\cite{hide3}}. Recently, Hayden \emph{et.al.}
{\cite{hide4}} gave an asymptotically secured data hiding scheme for
a large amount of quantum data in multipartite setting. However, we
consider here only hiding classical information in multipartite
quantum states. Multiparty data hiding is quite an interesting as
well as challenging job because of the strong security requirement.
Earlier, Eggeling \emph{et.al.} {\cite{hide2}} proposed a method for
hiding a classical bit in multipartite separable quantum states,
explicitly for $N=4$. In this work, a protocol is proposed for
hiding two classical bits rather than one cbit on activable bound
entangled states in multi-qubit systems. As a generalization of
Smolin state {\cite{smolin01}}, we found in any $2N$ qubit systems
for $N \geq 2$, there are always four orthogonal activable bound
entangled states \cite{bcbe}. The states are locally
indistinguishable. But, there are some limitations in providing a
hiding scheme. We investigate the possibility of hiding two bits of
classical information in those four states of $2N$ qubit system
shared between $2N$ number of distant parties.

Firstly, let us describe the class of activable bound entangled
states of multi-qubit system. The four qubit states are shared among
four distant parties by sharing equiprobable mixture of pairs of
Bell states taken in proper order.
\begin{equation}
\begin{array}{lcl}
\rho_4^{\pm} &=&\frac{1}{4}~ \{P[\Phi^+]\otimes
P[\Phi^\pm]+P[\Phi^-]\otimes P[\Phi^\mp]+ P[\Psi^+]\otimes
P[\Psi^\pm]\\&  &+P[\Psi^-]\otimes P[\Psi^\mp] \}\\

\sigma_4^{\pm} &=&\frac{1}{4}~ \{P[\Phi^+] \otimes P[\Psi^
\pm]+P[\Phi^-] \otimes P[\Psi^ \mp]+ P[\Psi^+]\otimes P[\Phi^
\pm]\\&  &+P[\Psi^-] \otimes P[\Phi^ \mp] \}  \label{abe}%
\end{array}
\end{equation}
where $ | \Phi^ \pm \rangle \equiv \frac{|00 \rangle\pm
|11\rangle}{\sqrt{2}}$ and $ |\Psi^ \pm \rangle \equiv \frac{|01
\rangle \pm |10 \rangle}{\sqrt{2}}$ are the Bell states, written in
their usual basis and $P[ \cdot ]$ represents projectors on those
states. The state $\rho_4^+$, known as Smolin state \cite{smolin01},
is used to perform various quantum information theoretic tasks like
secret key distillation, remote information concentration, etc.,
{\cite{use,use1}}. Afterwards it is generalized to a class of
activable bound entangled states in multiqubit systems
{\cite{bcbe}}. In any even number of qubit system starting from
four, there are exactly four states belonging to this class. A nice
Bell-correlation is seen in this class between the states of two
successive systems, that provides the generalization scheme. If we
denote the $2N$ qubit states as $\rho_{2N} ^{\pm}$, $\sigma_{2N}
^{\pm}$ then the next four states of $2N+2$ qubit system are given
by,
\begin{equation}
\begin{array}{lcl}
\rho_{2N+2} ^{\pm} &=& \frac{1}{4}~ \{ \rho_{2N}^{+} \otimes
P[\Phi^{\pm}]+\rho_{2N}^{-} \otimes P[\Phi^{\mp}]+ \sigma_{2N}^{+}
\otimes P[\Psi^{\pm}]\\&  &+\sigma_{2N}^{-}
\otimes P[\Psi^{\mp}] \}\\
\sigma_{2N+2} ^{\pm} &=& \frac{1}{4}~ \{ \rho_{2N}^{+} \otimes
P[\Psi^{\pm}]+\rho_{2N}^{-} \otimes P[\Psi^{\mp}]+ \sigma_{2N}^{+}
\otimes P[\Phi^{\pm}]\\&  &+\sigma_{2N}^{-} \otimes P[\Phi^{\mp}] \} \label{bcabe}%
\end{array}
\end{equation}

This correlation enables one to generate the whole class of states
from the four qubit states by a recursive process. Our aim is to
explore some special features of this class of states together with
some practical usefulness.

\textbf{Permutation Symmetry:} The whole class of states are
symmetric over all the parties concerned, i.e., the states remain
invariant under the interchange of any two parties. The four states
$\rho_{2N+2}^{\pm}$, $\sigma_{2N+2} ^{\pm}$ can be expressed as
$$\rho_{2N+2} ^{\pm} = \frac{1}{2^{2N}}~ \{~ \sum^{2^{2N}}_{i=1}
P[|\alpha^i_{2N+2}\rangle \pm
\overline{|\alpha^i_{2N+2}\rangle}]~\}$$
$$\sigma_{2N+2} ^{\pm} = \frac{1}{2^{2N}}~ \{ ~\sum^{2^{2N}}_{i=1} P[|\beta^i_{2N+2}\rangle\pm
\overline{|\beta^i_{2N+2}\rangle}]~\}$$ where
$\{|\alpha^k_{2N+2}\rangle, \overline{|\alpha^k_{2N+2}\rangle},
|\beta^j_{2N+2}\rangle, \overline{|\beta^j_{2N+2}\rangle}; k,
j=1,2,\cdots, 2^{2N}\}$ is the usual basis of $2N+2$ qubit system,
divided in four equal parts of $\frac{2^{2N+2}}{4}=2^{2N}$ number of
states, so that $|\alpha^k_{2N+2}\rangle$, $|\beta^j_{2N+2}\rangle$
can be expressed as
\begin{equation}
\begin{array}{lcl}
|\alpha^k_{2N+2}\rangle &=& |p^k_1\rangle \otimes |p^k_2\rangle
\otimes \cdots \otimes|p^k_{2N+2}\rangle~ ~\forall ~k=1,2,\cdots,
2^{2N}
\end{array}
\end{equation}where
$p^k_i \in \{0,1\}~ ~\forall ~i=1,2,\cdots,2N+2$ with $p^k_1=0 $ and
\begin{equation}
\begin{array}{lcl}
|\beta^j_{2N+2}\rangle &=& |q^j_1\rangle \otimes |q^j_2\rangle
\otimes \cdots \otimes|q^j_{2N+2}\rangle~ ~\forall ~j=1,2,\cdots,
2^{2N}
\end{array}
\end{equation} where
$q^j_i \in \{0,1\}~ ~\forall ~i=1,2,\cdots,2N+2$ with $q^j_1=0 $,
such that $$\sum_{i=1}^{2N+2} p^k_i=~0(mod~2)~,~\sum_{i=1}^{2N+2}
q^j_i =~1(mod~2)$$ (i.e. number of zero's in any
$\alpha^k_{2N+2}$($\beta^j_{2N+2}$) is even(odd)). The states
$\overline{|\alpha^k_{2N+2}\rangle}$ and
$\overline{|\beta^j_{2N+2}\rangle}$ are orthogonal to the states
$|\alpha^k_{2N+2}\rangle$ and $|\beta^j_{2N+2}\rangle$ respectively
for all possible values of $k$ and $j$. In the above form, if we
permute any two parties then all $|\alpha^k_{2N+2}\rangle$ for
$k=1,2,\cdots, 2^{2N}$, are interchanged within themselves and their
orthogonals $\overline{|\alpha^k_{2N+2}\rangle}$. Similarly for all
$|\beta^j_{2N+2}\rangle$'s for $j=1,2,\cdots, 2^{2N}$. This simple
property implies the permutation symmetry of all the four states in
$2N+2$ qubit system. In particular, the explicit form of the 4-qubit
states are,
\begin{equation}
\begin{array}{lcl}
\rho^{\pm}_4  ~  &=&\frac{1}{4}(P[0000\pm 1111]+ P[  0011\pm 1100]
~  + P[  0101\pm 1010] \\& & ~ ~ ~ + P[  0110\pm 1001]  )\\
\sigma^{\pm}_4  ~ & =&\frac{1}{4}(P[0001\pm 1110]+ P[  0010\pm 1101]
~ +P[ 0100\pm 1011]  \\& & ~ ~ ~  + P[  0111\pm 1000]
)\label{bcbe4sym}
\end{array}
\end{equation}where the permutation symmetry of the states over all
the concerned parties is very much clear.

\textbf{Orthogonality:} From Eq.(\ref{bcabe}) it is clear that the
four states of $2N+2$ qubit system are orthogonal to each other if
the $2N$ qubit states are so. Also from Eq.(\ref{abe}) we observe
that the four states $\rho_4^\pm~,~\sigma_4^\pm$ of four qubit
system are mutually orthogonal. Thus in a recursive way it provides
orthogonality of the four activable bound entangled states of any
even qubit systems starting from four.

\textbf{Local Indistinguishability:} The four states of $2N$ qubit
system, for $N \geq 2$ are locally indistinguishable. To prove it,
let us assume that for some value of $N\geq 2$, the four states
$\rho_{2N}^{\pm}$, $\sigma_{2N} ^{\pm}$ are locally distinguishable.
Now, consider the state,
$$\rho_{2N+2}^{+} = \frac{1}{4}~ \{ \rho_{2N}^{+} \otimes
P[\Phi^{+}]+\rho_{2N}^{-}
\otimes P[\Phi^{-}]+ \sigma_{2N}^{+} \otimes P[\Psi^{+}]\\
+\sigma_{2N}^{-} \otimes P[\Psi^{-}] \}$$ where the first $2N$
parties are $A_1, A_2, \ldots, A_{2N-1}, B_1$ and last two parties
are $A_{2N}, B_2$, i.e., the state is separable by construction in
$A_1 A_2 \ldots A_{2N-1} B_1 : A_{2N} B_2$ cut. Again, the state is
symmetric with respect to the interchange of any two parties, i.e.,
$\rho_{2N+2}^{+}$ has the same form if the first $2N$ parties are
$A_1, A_2, \ldots, A_{2N}$  and last two parties are $B_1, B_2$. If,
the four states $\rho_{2N}^{\pm}$, $\sigma_{2N} ^{\pm}$ are locally
distinguishable, then by LOCC only, $A_1, A_2, \ldots, A_{2N}$ are
able distinguish between the states $\rho_{2N}^{\pm}$, $\sigma_{2N}
^{\pm}$. The remaining state between $B_1$ and $B_2$, is then any
one of the Bell states correlated according as above, so that $A_1,
A_2, \ldots, A_{2N}$ are able to share a Bell state among $B_1$ and
$B_2$, which is impossible as initially there is no entanglement in
between $B_1$ and $B_2$. So, all the four states $\rho_{2N}^{\pm}$,
$\sigma_{2N}^{\pm}$ are locally indistinguishable for any $N\geq 2$.
Our protocol also suggest that the states are even probabilistically
indistinguishable for any $N\geq 2$, as it is impossible to share
any entanglement by LOCC between $B_1$ and $B_2$. Let us assume that
the four states are locally indistinguishable with some probability
$p>0$, then having shared the state $\rho_{2N+2}^{+}$ among the
$2N+2$ parties, any set of $2N$ parties may able to distinguish
their joint local system with that probability $1>p>0$ and
correspondingly they may share on average a portion of Bell state
among the other two parties. In this way it is possible to extract
on average some amount of entanglement by performing LOCC only. This
contradicts with the bound entangled nature of $\rho_{2N+2}^{+}$.
Thus the four states of $2N$ qubit system are even probabilistically
locally indistinguishable.

\textbf{Maximal Ignorance:} Ignorance of any one party(i.e., by
tracing out one qubit system) from any of the four states
$\rho_{2N+2} ^{\pm}$, $\sigma_{2N+2} ^{\pm}$ will results in the
state $\frac{1}{2^{2N+1}}I^{2N+1}$. To establish this result let us
first show that it is true for the 4-qubit states. The first of the
four qubit state is,
\begin{equation}
\begin{array}{lcl}
\rho_4^{+}&=&\frac{1}{4}~\{P[\Phi^+]\otimes
P[\Phi^+]+P[\Phi^-]\otimes P[\Phi^-]+ P[\Psi^+]\otimes P[\Psi^+]\\&
&+P[\Psi^-]\otimes P[\Psi^-]\}\\ &=& \frac{1}{4}~\{P[\frac{|00
\rangle+ |11\rangle}{\sqrt{2}}]\otimes P[\Phi^+]+P[\frac{|00
\rangle- |11\rangle}{\sqrt{2}}]\otimes P[\Phi^-]\\& &+ P[\frac{|01
\rangle+ |10\rangle}{\sqrt{2}}]\otimes P[\Psi^+]+P[\frac{|01 \rangle- |10\rangle}{\sqrt{2}}]\otimes P[\Psi^-]\}\\
&=& \frac{1}{8}~\{(P[|00 \rangle]+ P[|11\rangle])\otimes
(P[\Phi^+]+P[\Phi^-])+ (P[|01 \rangle]\\& &+ P[|10\rangle])\otimes
(P[\Psi^+]+P[\Psi^-])+(|00\rangle \langle11|+ |11 \rangle \langle
00|)\otimes(P[\Phi^+]\\& &-P[\Phi^-])+ (|01\rangle\langle10|+ |10
\rangle \langle 01|)\otimes(P[\Psi^+]-P[\Psi^-])\}
\end{array}
\end{equation}Thus tracing out first qubit system of the above state
we will obtain,
\begin{equation}
\begin{array}{lcl}
\rho'_3&=&\frac{1}{8}~\{(P[|0 \rangle]+ P[|1\rangle])\otimes
(P[\Phi^+]+P[\Phi^-])\\& &~~~~+ (P[|0\rangle]+
P[|1\rangle])\otimes (P[\Psi^+]+P[\Psi^-])\}\\
&=&\frac{1}{8}~(P[|0 \rangle]+
P[|1\rangle])\otimes(P[\Phi^+]+P[\Phi^-]+P[\Psi^+]+P[\Psi^-])\\
&=&\frac{1}{2^{3}}I\otimes I^{2}\\
&=&\frac{1}{2^{3}}I^{3}
\end{array}
\end{equation}

Similarly all the other three four qubit states have this property.
The next step is to prescribe a mathematical induction process to
prove this property for the whole class of states, taken into
consideration. The process will ensure that if the statement of the
property is true for the $2N$ qubit states then so also the $2N+2$
qubit states and thus proceeding from the 4 qubit states to the 6
qubit states, then from 6 qubit to 8 qubit and so on. For this
purpose, we assume that for some integer $N$, the four states
$\rho_{2N} ^{\pm}$, $\sigma_{2N} ^{\pm}$ have this property. Thus
tracing out the first qubit system of $\rho_{2N} ^{\pm}$,
$\sigma_{2N} ^{\pm}$, will results in $\frac{1}{2^{2N-1}}I^{2N-1}$.
Then applying the relation (\ref{bcabe}) we will show that, tracing
out the first qubit system of the state $\rho_{2N+2} ^{\pm}$ will
give $\frac{1}{2^{2N+1}}I^{2N+1}$. Taking trace over the first qubit
system of $\rho_{2N+2} ^{\pm}$ will produce,
\begin{equation}
\begin{array}{lcl}
\rho'_{2N+1}&=&~\frac{1}{4}\cdot\frac{1}{2^{2N-1}}I^{2N-1} ~\otimes
(P[\Phi^+]+P[\Phi^-]+P[\Psi^+]+P[\Psi^-])\\ &=&
\frac{1}{2^{2N+1}}~I^{2N-1}\otimes I^{2}\\ &=&
\frac{1}{2^{2N+1}}~I^{2N+1}
\end{array}
\end{equation}In a similar manner it can be shown that all the
four states $\rho_{2N+2} ^{\pm}$, $\sigma_{2N+2} ^{\pm}$ have this
property, if it holds for the $2N$ qubit states. Now, we already
found the result that the four qubit states have this property and
assuming the validity of this property for the four states of $2N$
qubit system, we find the property is also true for the $2N+2$ qubit
states. Thus through a recursive method we obtain, the property is
true for the whole class of states. As the states are symmetric over
permutation of all parties, thus tracing out any one party results
the same. This will also imply that the individual density matrices
of each party is a maximally mixed state, i.e., $\frac{1}{2}I$.

This class of states appears to be very suitable to construct a data
hiding protocol. Instead of finding two orthogonal mixed states to
hide one cbit of information, here we want to use all the four
orthogonal, highly symmetric mixed entangled states, to hide
classical bits. Our protocol is to hide two cbit of information
$b=0,1,2,3$ between $2N+2$ number of parties separated by distance,
by sharing the four states $\rho_{2N+2} ^{\pm}$, $\sigma_{2N+2}
^{\pm}$, for $N \geq 1$ among themselves. The hidden data is secured
against every possible LOCC among all the parties and against any
sort of quantum communication among $2N+1$ parties as the hidden
data can not retrieved perfectly, until and unless all the parties
remain separated or all of the $2N+2$ parties are dishonest.

\textbf{Security against LOCC:} The class of four states of $2N+2$
qubit system, for $N \geq 1$, used for sharing the data are locally
indistinguishable not only deterministically but also
probabilistically (shown earlier). So considering all the parties to
be dishonest, they can not even probabilistically recover the hidden
data perfectly by local operations on their subsystems and
communicating each other through some classical channel. However
imperfect knowledge of hidden data may be obtained by LOCC.

\textbf{Security against Global operation:} The data remains secure
under the action of any $2N+1$ number of dishonest distant parties,
who are allowed to make global operations, by joining in some labs
and make collective operation on their joint system. This follows
from the maximal ignorance property of the activable bound entangled
states, as ignorance of the system of the honest party (there should
be at least one such or otherwise the states are obviously globally
distinguishable as being orthogonal to each other) gives the reduced
density matrix of the others to be the maximally mixed state. Here
the quantum communication is allowable among a maximum number of
parties, i.e., $2N +1$.

It is interesting to note that in the above protocol we need only
one honest party, not allowed to communicate with the others through
some quantum channel. The hider may not be a part of the system. It
is also not necessary that the hider herself encrypt the bit in the
quantum state and thus knows the hidden data.

\textbf{Limitations regarding Inconclusive distinguishability:}
Although our protocol appears to be quite nice to maintain the
secrecy of the hidden data in a very stronger manner, but it has
some limitations. So far we have only considered perfect
distinguishability of the states. Precisely, it implies that we have
to discriminate the state supplied, from the whole set (here the set
of four states of $2N+2$ qubit system). However, it may be possible
to determine whether the given state belongs to some particular
subset of the whole set of states. i.e., although it is impossible
to distinguish perfectly the four states $\rho_{2N+2} ^{\pm}$,
$\sigma_{2N+2} ^{\pm}$, for $N \geq 1$ even with an arbitrarily
small probability by LOCC, but it is possible to determine by LOCC,
either the given state belongs to a subset containing any two of
$\rho_{2N+2} ^{+}, \rho_{2N+2} ^{-}, \sigma_{2N+2} ^{+}$ and
$\sigma_{2N+2} ^{-}$ or from other two. For example, if $\rho_{2N+2}
^{+}$ and $\rho_{2N+2} ^{-}$ are in a group and $\sigma_{2N+2}
^{+}$, $\sigma_{2N+2} ^{-}$ are in another group, then by measuring
on $\sigma_z$ basis in each party and checking only the parity (even
or odd number of zeros or ones), it is possible to discriminate any
state from the four $\rho_{2N+2} ^{\pm}, \sigma_{2N+2} ^{\pm}$, the
group it belongs. The basic fact of this set discrimination, taken
two together, is that the four states $\rho_{2N+2} ^{\pm},
\sigma_{2N+2} ^{\pm}$ are locally Pauli connected.

In conclusion we have obtained a class of highly symmetric activable
bound entangled states in any even number of parties that are
locally indistinguishable, if single copy of the states are given.
We have formulated a scheme to hide two bits of classical
information by sharing it among $2N+2$ number of parties, for any $N
\geq 1$. The advantage of our protocol is that the number of parties
can be extended in pairs up to any desired level keeping the
individual systems only with dimension two. The hidden information
can not be exactly revealed by any classical attack of the
corresponding parties and also against every quantum attack, as long
as one party remains honest. However, the hiding scheme has some
limitations from the viewpoint of set distinguishability. The states
are nice for practical preparation by sharing Bell mixtures among
distant parties. This class of locally Pauli connected but local
indistinguishable states with the power of activable boundness opens
a new direction in the
study of the relation between nonlocality and local distinguishability. \\

{\bf Acknowledgement.} The authors thank the referees for their
valuable suggestions and comments. They are also grateful to G. Kar
for useful discussions concerning the issue of this paper. I.C.
acknowledges CSIR, India for providing fellowship during this work.

\end{document}